\documentclass[journal]{IEEEtran}
\usepackage{graphicx,epsfig,amsmath,amssymb,bm}
\usepackage{amssymb,balance}
\usepackage{amsmath}
\usepackage{amsthm} 
\usepackage{amssymb}
\usepackage{algorithmicx}
\usepackage{algpseudocode}
\usepackage{cite}
\usepackage{url}
\newfont{\bbb}{msbm10 scaled 500}

\newfont{\bb}{msbm10 scaled 1100}

\newcommand{\RR}{\mbox{\bb R}}

\newcommand{\av}{{\bf a}}

\newcommand{\cv}{{\bf c}}

\newcommand{\ev}{{\bf e}}
\newcommand{\fv}{{\bf f}}

\newcommand{\nv}{{\bf n}}

\newcommand{\pv}{{\bf p}}

\newcommand{\rv}{{\bf r}}
\newcommand{\sv}{{\bf s}}

\newcommand{\xv}{{\bf x}}

\newcommand{\zv}{{\bf z}}

\newcommand{\Am}{{\bf A}}

\newcommand{\Dm}{{\bf D}}

\newcommand{\Hm}{{\bf H}}

\newcommand{\Wm}{{\bf W}}

\newcommand{\thetav}{\hbox{\boldmath$\theta$}}

\usepackage{bbm}
\usepackage{subcaption}
\usepackage{tabularx}
\usepackage{multirow}
\usepackage{verbatim}
\usepackage[ruled,linesnumbered]{algorithm2e}

\usepackage{color}
\usepackage{tikz}
\usepackage[acronym]{glossaries}
\usepackage[capitalise,noabbrev]{cleveref}
\usepackage{makecell}
\newcounter{savefig}

 \def\b0{\mbox{\boldmath $0$}}

\usepackage{booktabs}
\usepackage{bbding}
\usepackage{pifont}

\def\argmin{\operatornamewithlimits{arg\,min}}

\newcommand{\beqa}{\begin{eqnarray}}
\newcommand{\eeqa}{\end{eqnarray}}

\newcommand{\sub}[1]{{\color{red} [Sub: #1]}}

\ifCLASSINFOpdf

\else

\fi

\newglossary[slg]{symbolslist}{syi}{syg}{List of Symbols}

\begin{document}

\title{Survey of Moving Target Defense in Power Grids: Design 
Principles, Tradeoffs, and Future Directions}

\author{
    Subhash~Lakshminarayana,~\IEEEmembership{Senior Member, IEEE}, 
    Yexiang Chen,
    Charalambos Konstantinou,~\IEEEmembership{Senior Member, IEEE}, 
    Daisuke Mashima,~\IEEEmembership{Member, IEEE}, 
    and Anurag K. Srivastava,~\IEEEmembership{Fellow, IEEE} \vspace{-0.2in}
    
}

\IEEEaftertitletext{\vspace{-1.2\baselineskip}}
\maketitle

\begin{abstract}
Moving target defense (MTD) in power grids is an emerging defense technique that has gained prominence in the recent past. It aims to solve the long-standing problem of securing the power grid against stealthy attacks. The key idea behind MTD is to introduce periodic/event-triggered controlled changes to the power grid's SCADA network/physical plant, thereby invalidating the knowledge attackers use for crafting stealthy attacks. In this paper, we provide a comprehensive overview of this topic and classify the different ways in which MTD is implemented in power grids. We further introduce the guiding principles behind the design of MTD, key performance metrics, and the associated trade-offs in MTD and identify the future development of MTD for power grid security. 

\end{abstract}


\IEEEpeerreviewmaketitle
\vspace{-3mm}
\section{Introduction} 
\vspace{-1mm}

Modern power grid monitoring relies on state estimation (SE) through Supervisory Control and Data Acquisition (SCADA), which is essential for optimal power flow (OPF) and contingency management. While integrating information and communication technology (ICT) enhances efficiency, it also increases the risk of cyber-physical attacks. SE is particularly vulnerable to attacks where measurements/control signals can be compromised. Stealthy attacks, such as false data injection (FDI), can bypass detection mechanisms like bad data detectors (BDD), leading to significant damage and economic loss. Securing the grid against these attacks remains a critical challenge \cite{MashimeNOW2023}.

To address cyber-physical threats to power grids, several defense approaches have been proposed. Firstly, security can be enhanced through hardware updates (e.g., via additional protective measures or upgrading the security features of existing devices). Enhancements in sensor security can be achieved by incorporating more data integrity checks and independent verification of system states using Phasor Measurement Units (PMUs) \cite{Bobba2010, KimSensor2013, YangAggregation2013}. Alternatively, grid security can be enhanced by preventing unauthorized access, e.g., via authentication technologies able to prevent attackers from injecting malicious commands or data \cite{esiner2022lomos, tefek2022caching}. 
Implementing security upgrades in SCADA can be costly, and solutions like those involving PMUs, may still be vulnerable to GPS spoofing attacks \cite{konstantinou2017gps}.

Beyond hardware and encryption upgrades, security can be improved through advancements in detection models and algorithms, including traditional detectors like bad data detection (BDD) \cite{abur2004power}, Kalman filter-based detectors \cite{simon2010kalman}, and advanced methods such as CUSUM-based detectors \cite{huang2011defending}. However, attackers with detailed knowledge of grid topology can bypass these detectors \cite{liu2011false}. Machine learning (ML), particularly deep learning (DL) methods like deep neural networks (DNNs), have gained interest for detecting attacks that evade BDD systems \cite{ozay2015machine, he2017real, ZhangSemi, lakshminarayana2022data, mohammadpourfard2021attack}. Despite their effectiveness, ML techniques can be susceptible to adversarial attacks \cite{goodfellow2015explaining, huang2022adversarial, AFDIA2021}, limiting their reliability in practical applications.


Despite significant efforts, static defense mechanisms remain vulnerable to advanced, persistent attacks. Attackers often gain the upper hand by conducting prolonged reconnaissance, as seen in the 2015 BlackEnergy attack on Ukraine's power grid, where the adversary spent months gathering operational details before executing the attack \cite{case2016analysis}. This issue is particularly evident in FDI attacks, where attackers can bypass BDD by monitoring grid measurements over time \cite{chin2017blind, LakshRMT2021}. 



\vspace{-3mm}
\subsection{Introduction to Moving Target Defence in Power Grids}
\vspace{-1mm}

\begin{figure}[t]
\centering
\includegraphics[width=\columnwidth]{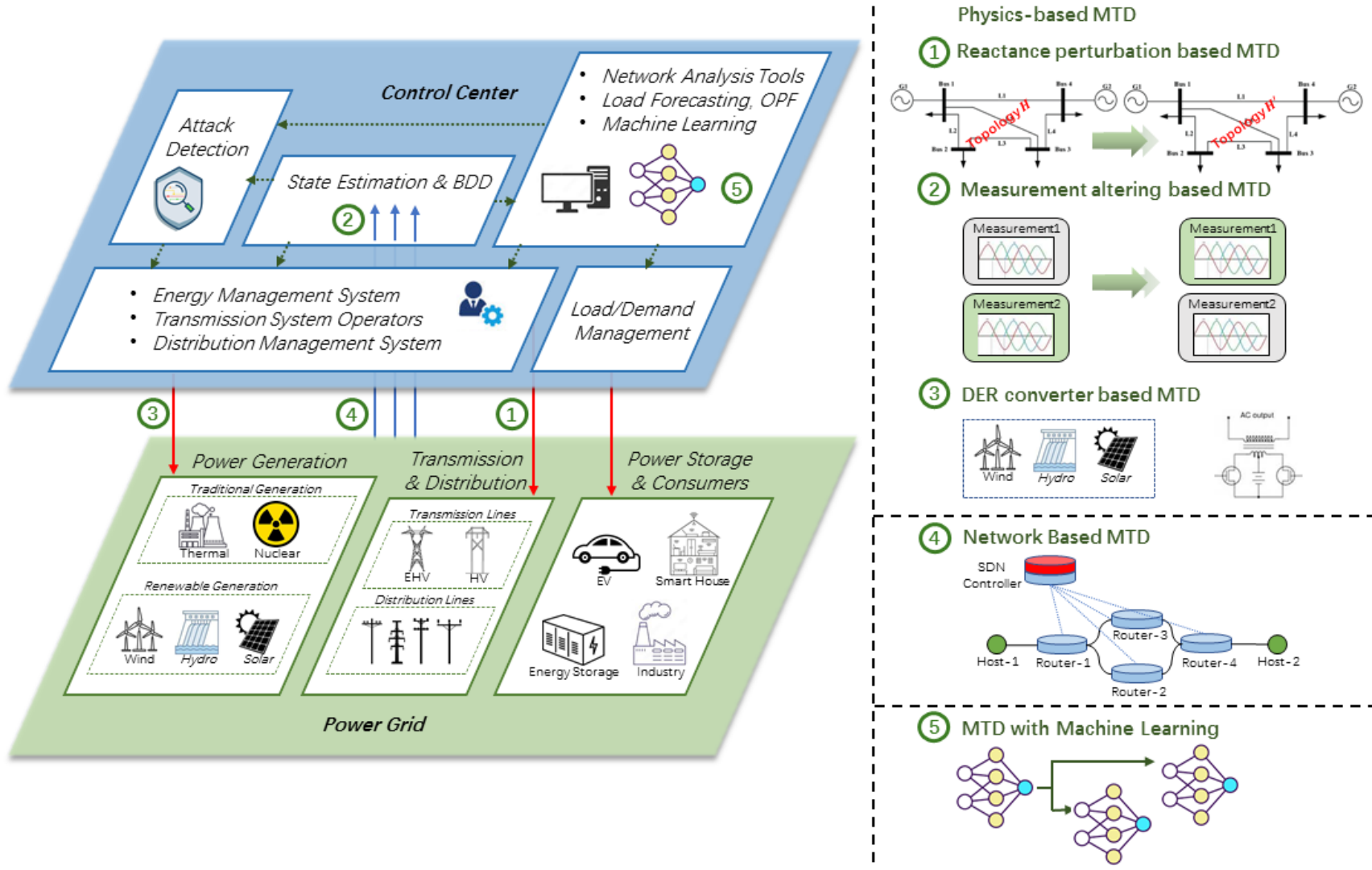}
\vspace{-2mm}
\caption{Pictorial depiction of MTD strategies within a cyber-physical power system.}
\label{fig:Pictorial Depiction of MTD in Power Grids}
\vspace{-5mm}
\end{figure}

Moving target defense (MTD) has recently emerged as a proactive strategy to thwart stealthy attacks in power grids by increasing the cost and complexity for attackers. MTD disrupts attackers'  reconnaissance efforts by continuously reconfiguring system settings (e.g., SCADA network or physical plant configurations), invalidating their knowledge of the system. 
The concept of MTD has roots in several areas of computer security, including fault tolerance \cite{chen1978n, avizienis1985n}, reconfigurable computing \cite{compton2002reconfigurable, baran1989network}, and bio-inspired cybersecurity \cite{lala2009monoculture}. MTD enhances system resilience by exposing or preventing attacks based on outdated information. 
MTD has been widely applied in many fields such as cloud computing \cite{zhang2012incentive}, control systems \cite{griffioen2020moving}, automotive systems \cite{potteiger2022tutorial}, power systems \cite{LakshGT2021} and artificial intelligence \cite{Morphence2021}. A pictorial depiction of MTD in power grids is shown in Fig.~\ref{fig:Pictorial Depiction of MTD in Power Grids}. 

Unlike IT security techniques requiring major upgrades, such as encryption-enabled PLCs or remote attestation, MTD enhances system security by re configuring the existing devices in the grid. In power systems, MTD can be deployed as a standalone defense or integrated with other strategies, like data-driven detectors \cite{ozay2015machine, he2017real}. MTD is effective for transmission \cite{LakshTPWRS2021, FeiMTD2021}, distribution systems \cite{JhalaDistibution2021, CuiHidden2021, Liu2018}, and microgrids \cite{LiuMCMTD2021, GiraldoMTDMicrogrid}, making it a likely candidate for future adoption by power grid operators.

\vspace{-3mm}
\subsection{{Difference From Existing Works}}
\vspace{-1mm}
In recent years, several survey papers \cite{cai2016moving, lei2018moving, ward2018survey, zheng2019survey, sengupta2020survey, navas2020mtd, MTDSurvey2020, Zhang2021survey, sun2023survey, Abdi2024survey}, have emerged on MTD, covering a wide range of techniques, categorization principles, evaluation criteria, and design strategies. However, the primary focus of these works is on applying MTD for network security, Internet-of-Things (IoT), and general cyber-physical systems (CPS). For instance,  \cite{ward2018survey} classifies network-based MTD into five categories based on how the MTD is implemented, i.e., dynamic data, dynamic software, dynamic runtime environment, dynamic platforms, and dynamic networks. Other works, such as \cite{cai2016moving}, investigate both network-based and physics-based MTD, categorizing technologies according to their application scenarios. \cite{lei2018moving} categorizes MTD based on theory, techniques, and purpose, while \cite{zheng2019survey} organizes MTD according to its architecture. In \cite{sengupta2020survey}, the authors categorize MTD based on designing key movements. Notably,  \cite{farris2015quantification} and \cite{sengupta2020survey} focus on quantifying and evaluating the effectiveness and cost of MTD strategies. Furthermore, \cite{MTDSurvey2020} provide a comprehensive review and investigate measurement methods, metrics, common attackers countermeasures by MTD, distinctions between MTD and alternative defense mechanisms, and discussions comparing MTD with other defense strategies. However, none of these surveys specifically address the application of MTD in the context of power grid security, which has unique challenges and design principles.

In the area of power grid security, particularly focusing on false data injection (FDI) attacks, \cite{Deng2017survey} provides a comprehensive survey of attack mechanisms and defense strategies, including protection of meter measurements and PMU-based security. \cite{Zhang2021survey} summarizes stealthy attacks in power grids and explores cyber-physical defense strategies, offering a brief overview of MTD, though limited to transmission grids and reactance perturbation-based MTD. Other surveys in power grid literature, such as \cite{sayghe2020survey, Abdi2024survey}, do not thoroughly address MTD's application in enhancing grid security. While \cite{Abdi2024survey} discusses the use of DL for proactive defense, it lacks a clear description of network-based and physics-based MTD in power grids. Our survey fills this gap by offering a detailed categorization and analysis of both, along with a comprehensive examination of the underlying mechanisms and motivations of MTD in power grids.

\vspace{-3mm}
\subsection{Paper Contributions} \label{sec: Paper Contributions}
\vspace{-1mm}
This paper is the first to provide a comprehensive survey of MTD in power grids encompassing several aspects. The key contributions of this survey paper are as follows.
\begin{itemize}
    \item Categorizing the implementation of MTD in power grids into (i) physics-based MTD, (ii) network-based MTD, (iii) deception-based MTD, and (iv) MTD with ML. 
    \item Enumerating the criteria and metrics to evaluate the performance of MTD in power grids.
    \item A detailed analysis of physics-based MTD in power grids to answer the fundamental questions on what, when and how to move for effective MTD implementation.  
    \item Detailed analysis of the implementation of MTD in less-explored scenarios such as distribution systems/microgrids, network-based MTD and the combination of MTD with other ML techniques to enhance its effectiveness.
    \item Finally, a description of the open research directions for MTD in power grids.
\end{itemize}

The remainder of the paper is organized as follows: Section \ref{Overview of Moving Target Defense in Power Grids} provides an overview of MTD in power grids. Sections \ref{Physics-Based MTD} and \ref{Network-based MTD} introduce physics-based and  network-based MTD strategies, respectively. Section \ref{MTD with machine learning} discusses the integration of MTD with ML techniques. Finally, Section \ref{Conclusion and future work} concludes the survey and outlines directions for future research.

\vspace{-3mm}
\section{Overview of MTD in Power Grids} \label{Overview of Moving Target Defense in Power Grids}
\vspace{-1mm}

MTD's core concept involves dynamically reconfiguring system or model parameters to invalidate attackers' knowledge and thwart stealthy attacks. In this section, we examine key MTD techniques used in power grids, detailing their categories, application scenarios, required devices, and associated costs. Table~\ref{tbl:Power grid model and MTD implementation} provides an overview, followed by an in-depth explanation of these techniques.

\vspace{-3mm}
\subsection{Assumptions on the Attacker's Capability}
Other than the commonly considered assumption that the attacker can compromise multiple nodes within the grid to launch sophisticated attacks such as FDI, the typical assumptions about the attacker's capability when integrating MTD in power grids include that the attacker has significant knowledge of the power grid's topology and operational parameters, obtained through extensive reconnaissance \cite{RahmanMTD2014}. This knowledge may be acquired through methods such as topology leaking attacks \cite{Markwood2017}. Attackers are also assumed to need time to gather knowledge, as methods like topology attacks rely on historical data. Moreover, they are considered capable of adapting strategies based on system changes, using advanced tools to analyze MTD patterns, monitor in real-time, and potentially exploit temporary vulnerabilities during MTD transitions. These assumptions ensure MTD strategies are robust against highly knowledgeable and adaptive adversaries, addressing worst-case scenarios. 

\vspace{-3mm}
\subsection{MTD Implementation in Power Grids} \label{sec: MTD Implementation in Power Grids}
\vspace{-1mm}
\noindent $\bullet$ {\bf Physics-Based MTD:}  
Physics-based MTD entails adjusting the settings or parameters of physical devices of the power grid. This may involve actions such as altering transmission line reactances, randomizing the set of sensor measurements used for SE, or adjusting the control signal of distributed energy sources (DER) converters. Such changes are aimed at invalidating the attackers' knowledge of the underlying physical process, making it more challenging for them to craft attacks that can bypass physics-based attack detectors. 
A major advantage of this approach is that it leverages existing devices in the physical network, such as Distributed Flexible AC Transmission Systems (D-FACTS). 
Consequently, it does not require significant hardware updates, such as deploying encryption-enabled devices. However, the MTD perturbations can typically lead the system away from its optimal settings (e.g., changes to line reactance settings). Therefore, the implementation of physics-based MTD results in penalties, such as increased OPF costs or power losses. Next, we enlist the commonly considered approaches.

 \emph{Reactance perturbation based MTD:} 
        This approach involves perturbing the transmission line reactance using D-FACTS devices. The motivation is the need for attackers to obtain the knowledge of the power grid topology and the corresponding measurement matrix in order to launch BDD-bypassing FDI attacks \cite{liu2011false}. Thus, reactance perturbations will invalidate this knowledge and make the attacks detectable. However, reactance perturbation-based MTD results in an unavoidable increase in power losses \cite{liu2020optimal} and OPF cost \cite{LakshTPWRS2021} and may also raise voltage stability issues \cite{CuiHidden2021}. Therefore, the perturbation must be carefully designed to minimize its negative effects while achieving security goals.
        
   \emph{ Measurement-Altering MTD:} This approach involves dynamically changing the subset of measurements used for SE and control decisions. Since full observability in a power system can be achieved with only a subset of sensor measurements, operators have the flexibility to select which measurements to use. This randomization prevents attackers from knowing which sensors are being used for SE, making it difficult to target specific sensors for their attacks \cite{RahmanMTD2014, hu2023controlled}.
    
    \emph{ DER converter-based MTD:} This strategy involves manipulating the control signal of the DER converter to expose covert attacks. For instance, \cite{LiuMCMTD2021} implemented MTD in a microgrid by adjusting the primary control gain (PCG) of the DER converter. This technique is then utilized to strengthen an unknown input observer (UIO)-based detector. Similarly, \cite{JhalaDistibution2021} implemented MTD by employing inverter-based DERs to generate perturbation signals in the voltage of distributed systems. The motivation for using MTD with DER converters is to prevent attackers from exploiting system vulnerabilities, including zero-day vulnerabilities, to launch attacks.

\begin{table*}[!h]
\centering
    \begin{tabular}{|l|p{1.1cm}|l|l|p{3cm}|p{5cm}|}
        \hline
        MTD category & Reference & \multicolumn{2}{l|}{Application Scenario} & MTD Implementation & Penalty \& Cost \\
        \hline
        \multirow{10}{*}{Physics-based} & \makecell[l]{\cite{RahmanMTD2014, Zhang2020}} & \multirow{5}{*}{Trans.} & DC & \makecell[l]{Perturbs D-FACTS devices \\ + Perturbs measurements} & \multirow{10}{*}{\parbox{\linewidth}{Increase OPF costs, power loss and cause voltage instability}} \\ 
        \cline{2-2}
        \cline{4-5}
         & \makecell[l]{\cite{Morrow2012,Davis2012}} & & DC & \multirow{5}{*}{Perturbs D-FACTS devices} & \\
        \cline{2-2}
        \cline{4-4}
         & \makecell[l]{\cite{xu2021robust, liu2020interior} \\ \cite{Liu2022}} & & AC &  & \\
        \cline{2-2}
        \cline{4-4}
         & \makecell[l]{\cite{LakshTPWRS2021, liu2020optimal} \\ \cite{TianHiddenMTD2019, LiuBo2021} \\ \cite{LiuFDI2019} } & & DC +  AC &  & \\
         \cline{2-4}
         & \makecell[l]{\cite{CuiHidden2021}} & \multirow{3}{*}{Distr.} & 3-phase AC&  & \\
        \cline{2-2}
        \cline{4-4}
         & \makecell[l]{\cite{Liu2018}}  & & AC & & \\ 
        \cline{2-2}
        \cline{4-5}
         & \makecell[l]{\cite{JhalaDistibution2021}} & & AC & \multirow{2}{*}{Perturbs converter signal} & \\ 
         \cline{2-4}
         & \makecell[l]{\cite{LiuMCMTD2021}} & \multirow{2}{*}{Microgrid} & DC &  &\\ 
        \cline{2-2}
        \cline{4-5}
         & \makecell[l]{\cite{GiraldoMTDMicrogrid}} & & AC & \makecell[l]{Replication signal \\ + Path selection} & \\
        \hline
        Network-based & \makecell[l]{\cite{pappa2017moving, ulrich2017symmetric} \\ \cite{MTDSDN, lin2020defrec} \\ \cite{yang2020decied}} &    \multicolumn{2}{c|}{ \makecell[l]{SCADA \\ Communication system} } & Alters the path of communication signal & Increase the packet drop percentages, reduce throughput and cause delay\\
        \hline
        MTD with ML & \makecell[l]{\cite{fMTD2019, sengupta2019mtdeep} \\ \cite{Morphence2021}} & \multicolumn{2}{c|}{ \makecell[l]{Image Classification} } & Perturbs ML decision functions & Computational resources and time for retraining. Increase in FP rate. \\
        \hline
    \end{tabular}
    \caption{MTD categories and application scenarios.}
    \vspace{-5mm}
	\label{tbl:Power grid model and MTD implementation}
\end{table*}

     $\bullet$ {\bf Network-Based MTD:} 
    Network-based MTD involves reconfiguring network devices, with techniques like IP-hopping or dynamically altering SCADA communication routes within software-defined networks \cite{pappa2017moving, ulrich2017symmetric, MTDSDN}. This strategy aims to prevent attackers from identifying targets, such as communication paths. However, penalties include increased packet loss, reduced throughput, and delays. Current literature mainly addresses basic attacks like denial-of-service (DoS) while lacking countermeasures for more sophisticated threats like deception, FDI, or replay attacks. Additionally, the design principles of network-based MTD often lack in-depth analysis. 
      
 $\bullet$ {\bf {MTD with Deception Technology}}
Deception technologies, such as honeypot and decoy networks, are cybersecurity solutions that aim at confusing attackers by deploying a number of dummy, often virtual, devices that appear and behave like real devices. Deception technologies for smart grids have been explored in the literature \cite{lin2020defrec, yang2020decied}. Deception technologies can implement MTD in the network layer by dynamically changing network addresses as well as topologies to make it difficult to find real system configurations. Moreover, such decoy devices, which work as sensors in the smart grid, can further send crafted measurements based on physics-based MTD, which are elaborated on later in this paper, to present a fake view of the physical power system to mislead the attack tactics, without modifying real power grid status and/or device configurations. In this sense, deception/decoy devices can be a less intrusive platform for deploying MTD.
    
$\bullet$  {\bf MTD with ML:} The integration of ML into MTD can be understood from two perspectives. Firstly, ML technologies are incorporated to enhance existing MTD approaches, such as physics-based MTD. The motivation behind this integration is that ML has the potential to boost efficiency and reduce costs associated with MTD. An example of this is the event-triggered MTD \cite{XuEventTriggered2023}, which utilizes ML to decrease the frequency of MTD updates. Secondly, there is ML-based MTD, which entails directly applying MTD mechanisms to ML functions to reinforce ML-based detection systems. The motivation behind this approach is rooted in the vulnerability posed by attackers obtaining the parameters of ML-based detectors, enabling them to launch adversarial attacks aimed at bypassing detection \cite{AFDIA2021}. ML-based MTD involves dynamically generating multiple new ML-based detectors and collaboratively making detections \cite{fMTD2019, Morphence2021}. This strategy relies on designing different ML models such that the transferability of adversarial attacks across different ML-based detectors is reduced. However, ML-based MTD may require more computational resources, memory, and time for retraining, potentially leading to an increased false positive rate on legitimate measurements.

\vspace{-3mm}
\subsection{Criteria and Metrics to Evaluate MTD's Performance} \label{sec: Criteria and Metrics to Evaluate MTD's Performance}
Next, we elaborate on common criteria and metrics to evaluate MTD's performance. The three key criteria include (i) attack detection effectiveness, (ii) capital and operational cost to implement MTD, and (iii) hiddenness of MTD. 

\subsubsection{The Attack Detection Effectiveness of MTD}
MTD's primary objective is to detect stealthy attacks, so this is the most important criterion to assess MTD's performance. The effectiveness can be measured based on two notions. 

\noindent $\bullet$ {\bf MTD Admitting No Undetectable Attacks:} The first notion of MTD's effectiveness is to prevent an attacker's ability to craft undetectable attacks by invalidating their knowledge. In other words, following the system reconfiguration, no attack crafted with the outdated system knowledge goes undetected. In the context of FDI  against SE, this implies that there must be no attacks whose detection probability is as low as the false alarm rate. This notion is often referred to in research literature as \emph{complete} of MTD \cite{LiuMTD2018}. 
However, achieving complete MTD is typically challenging due to practical constraints. Incomplete MTD, therefore, focuses on maximizing the probability of attack detection under practical constraints.

\noindent $\bullet$ {\bf MTD Ensuring High Attack Detection Rate:} The completeness of MTD, however, does not guarantee a high detection rate (also known as worst-case detection rate) for attacks \cite{LakshTPWRS2021, xu2021robust}. For instance, in the context of FDI attacks, although MTD may prevent the BDD triggering rate on stealthy attacks from being as low as the FP rate, it does not ensure a high detection probability. This is especially true in scenarios with noisy measurements \cite{xu2021robust}. Thus, effective MTD must ensure that the attacks are also detected with a high probability.

\subsubsection{The Cost of MTD}

{The cost and penalty of implementing MTD in power systems involve several aspects and can be different according to the categories of MTD. Generally, the cost of MTD comprises
two components. (i) \emph{Investment cost:} This cost is associated with the necessary hardware and software upgrades to achieve MTD strategies. For physics-based MTD, the investment cost involves installing additional D-FACTS, measurement units, and converters or upgrading existing ones to support MTD. For network-based MTD, the investment cost includes modifications or upgrades to routers, switches, controllers, and SCADA systems to support the dynamic routing enabled by SDN. (ii) \emph{Operation Cost:} This ongoing expense is typically associated with penalties incurred due to the MTD perturbations. For example, in the absence of physics-based MTD, the system's configuration (such as generator set points, line reactances, etc.) is set to minimize the system's operational cost. The perturbations due to physics-based MTD will alter these setpoints and, hence, incur a cost. This cost can be quantified as the increase in OPF cost \cite{LakshTPWRS2021} or the increase in real power loss of system \cite{LiuMTD2018}. For network-based MTD, the operational cost can be the increase in packet drop rates, decreased throughput, and increased delay \cite{MTDSDN}.

}
 
\subsubsection{The Hiddenness of MTD} 

Another criterion for MTD design is the notion of \emph{hiddenness}, which refers to hiding the activation of MTD from the attacker. The rationale is that if attackers detect the existence of MTD (e.g., through observing differences in measurement residual \cite{TianHiddenMTD2019, LiuBo2021, Liu2022, CuiHidden2021}, using an MTD-confirming detector \cite{Zhang2020}, or utilizing ML techniques \cite{FeiMTD2021}), they will adopt more cautious and sophisticated attack strategies. For example, attackers may increase reconnaissance efforts to understand MTD patterns and launch adaptive attacks, such as parameter confirming first FDI attacks \cite{TianHiddenMTD2019}. Therefore, enhancing the hiddenness of MTD from attackers can further improve its defense effectiveness.

In the rest of the paper, we will present a detailed survey of the MTD techniques stated in Section~\ref{sec: MTD Implementation in Power Grids} and elaborate on how they are designed to achieve the MTD's performance criteria in Section~\ref{sec: Criteria and Metrics to Evaluate MTD's Performance}.

\vspace{-3mm}
\section{{Physics-Based MTD}} \label{Physics-Based MTD}
Physics-based MTD relies upon changing the parameters of the power grid physical network. In this section, we review the various applications of physics-based MTD.  
\vspace{-3mm}
\subsection{{Reactance Perturbation-Based MTD Against FDI Attacks}}
In power system applications, the reactance perturbation-based MTD has been extensively studied. An FDI attack vector that lies within the column space of the power system's measurement matrix will remain undetected by the grid's BDD \cite{liu2011false}. It has been shown that a sophisticated attacker can learn the knowledge required to compute the grid's measurement matrix and, hence, construct undetectable attacks by continuous reconnaissance. For instance, this can be done by monitoring the power grid measurements over time \cite{chin2017blind, LakshRMT2021}.
Reactance-perturbation based MTD invalidates the attacker's knowledge of the power grid topology and the corresponding measurement matrix. 
This section provides a comprehensive analysis of reactnace-perturbation based MTD. Existing literature suggests implementing reactance perturbation-based MTD using D-FACTS devices, such as Distributed Static Series Compensators (DSSC) and Distributed Series Reactors (DSR) \cite{DFACTS2005}. These devices, which attach directly to transmission lines, dynamically control line impedance. Initially designed to enhance grid stability by providing rapid voltage, inductive, and reactive power support, they are also used for contingency management and power loss minimization \cite{DFACTS2005, RogersDFACTS2008}. 



\subsubsection{Designing MTD Reactance-Perturbations}
The design of MTD perturbations primarily involves D-FACTS operation, i.e., determining which D-FACTS device to operate and the magnitude of reactance perturbations. 
The initial work on this topic \cite{Morrow2012, Davis2012, RahmanMTD2014} proposed MTD design by introducing reactance perturbations whose magnitude is chosen randomly, which is unknown to the attacker. However, these approaches cannot offer performance guarantees in terms of attack detection rates. Furthermore, the impact of these perturbations on the power system's operation was not quantified. In the following, we conduct a detailed analysis of MTD design considering its effectiveness, cost and hiddenness.




\begin{figure}[!t]
\centering
\includegraphics[width=0.4\textwidth]{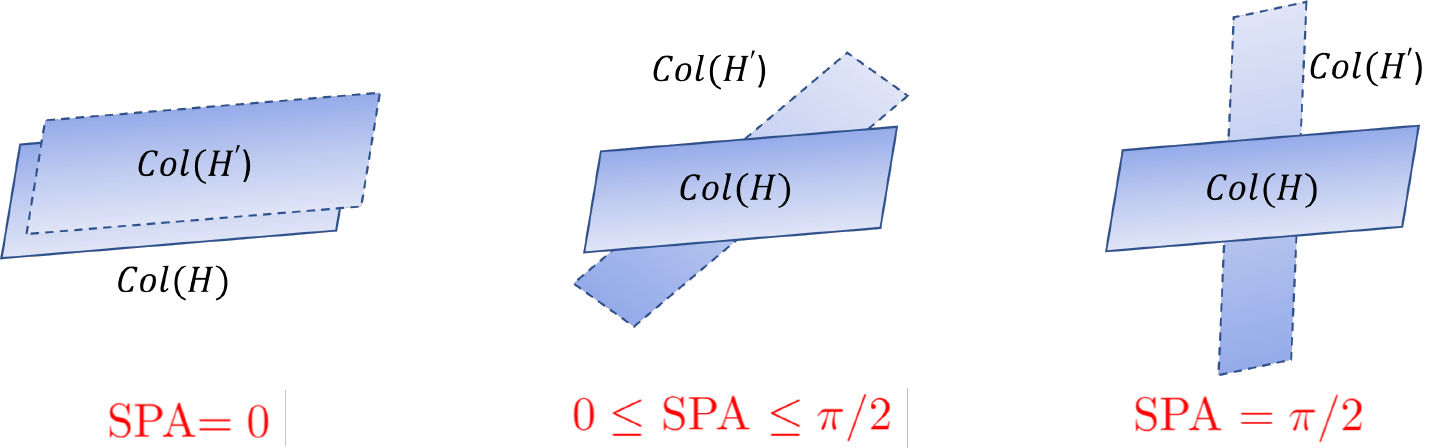}
\vspace{-1mm}
\caption{Smallest principle angle for assessing separation of column space and MTD effectiveness.}
\label{fig:MTD_SPA}
\vspace{-5mm}
\end{figure}

{\bf Attack Detection Effectiveness:} Effectiveness refers to the ability of MTD to detect FDI attacks that are crafted based on the outdated knowledge of the system (i.e., before the MTD perturbations).
Depending on its detection performance, an MTD scheme can be characterized either as a \emph{complete MTD} or an \emph{incomplete MTD} \cite{LiuMTD2018, TianHiddenMTD2019}. Complete MTD ensures that an attacker cannot construct an undetectable FDI attack using outdated knowledge of the power grid's measurement matrix.  \cite{LiuMTD2018} show that complete MTD can be achieved if the composite matrix, formed by the measurement matrix before and after MTD activation, is of full column rank. Similarly, \cite{LakshTPWRS2021} proposes that complete MTD can be achieved if the column spaces of the power grid measurement matrices, before and after MTD activation, are orthogonal. The two conditions can be shown to be equivalent. 

Achieving complete MTD is often challenging in real-world power systems due to constraints on the system topology and meter deployment. For instance, \cite{LiuMTD2018} shows that complete MTD requires the number of branches in the power system to be at least twice the number of system states, and the number of meters must be no less than twice the number of system states. Furthermore, \cite{ZhangMTD2020} shows that in order to achieve complete MTD, each bus must be connected by at least two branches, and the number of perturbed branches must not be fewer than the number of system states, and the perturbed branches should cover all buses. Most practical power systems lack a sufficient number of transmission lines, meters, or D-FACTS device deployment to achieve the aforementioned conditions for achieving complete MTD. Thus, in a practical setup, only the implementation of incomplete MTD is feasible in most cases. Various theoretical metrics have been designed to maximize the detection capability of incomplete MTD schemes within the power network constraints (i.e., topology, meter deployment, number of D-FACTS, etc.). 





In \cite{LakshTPWRS2021}, the smallest principal angle (SPA) was introduced to quantify the relationship between measurement matrices before and after MTD, as shown in Fig. \ref{fig:MTD_SPA}. Higher SPA values indicate more effective MTD for attack detection. Similarly, \cite{LiuMTD2018} shows that higher composite matrix rank improves MTD effectiveness. \cite{ZhangMTD2020} introduced the concept of stealthy attack space, aiming to minimize its dimension while maximizing bus coverage by MTD. \cite{zhang2021strategic} demonstrated that coordinated perturbations within an MTD cycle can enhance effectiveness. Comparing SPA and rank-based metrics, \cite{xu2021robust} found that SPA provides robust performance in noisy environments, while rank-based MTD is more effective in noiseless scenarios but less reliable with noise. 

All of the works above are based on DC power flow. \cite{Liu2022} explicitly considered MTD design using the AC power flow model. They proposed a measurement residual-based criteria. Specifically, they derived explicit approximations of measurement residuals to quantify the effectiveness of attack detection. By utilizing the projection matrix, they transform the problem of maximizing detection effectiveness into maximizing the lower bound of the approximated residual, which addresses the issue related to matrix inversion.

{\bf Operational Cost:} Operational cost is a key factor in MTD design, typically quantified by increases in OPF costs \cite{LakshTPWRS2021} or real power losses \cite{LiuMTD2018} due to MTD perturbations. \cite{LakshTPWRS2021} highlights the trade-off between MTD effectiveness and cost, proposing an optimization function that balances these by treating reactance perturbation as the independent variable. Similarly, \cite{LiuMTD2018} presents a weighted optimization problem minimizing power losses while maximizing MTD effectiveness. \cite{liu2020optimal} suggests that D-FACTS deployment drives MTD effectiveness, while D-FACTS perturbations primarily impact operational costs, leading to strategies focused on reducing these costs, such as minimizing power loss \cite{liu2020optimal} or OPF costs \cite{liu2020interior}. \cite{LakshGT2021} applies game theory to further reduce operational costs by focusing MTD on protecting critical system assets. 



{\bf MTD's Hiddenness:} Existing literature also addresses MTD's hiddenness. \cite{TianHiddenMTD2019} shows that MTD can remain hidden if power flows remain unchanged after perturbing D-FACTS devices, though it cannot fully achieve attack detection. They compute system states post-MTD and minimize power flow differences to maintain hiddenness. However, \cite{Zhang2020} notes that MTD may lose hiddenness over time due to power flow changes, proposing additional protection for measurements beyond the power grid's spanning tree to maintain hiddenness. \cite{LiuBo2021} suggests that MTD can stay hidden if power flows on non-perturbed lines remain unchanged, optimizing D-FACTS line susceptance while using hiddenness as a constraint. In a three-phase unbalanced system, \cite{CuiHidden2021} introduces 'deep-hidden MTD,' ensuring hiddenness by concealing both self and mutual reactance of transmission lines, integrating branch and injection power phasor measurements to maintain stability. 

In conclusion, D-FACTS operation strategies involve balancing balance MTD's effectiveness, operational cost, and hiddenness. This balance is typically achieved through optimization formulations. We summarize the optimization models proposed in the existing literature, detailing their objectives, constraints, and characteristics in Table~\ref{tbl:MTD operation}.

\subsubsection{D-FACTS Deployment Strategy} 
The discussion in the previous subsection only focuses on designing the reactance perturbations for a given D-FACTS deployment. In this subsection, we review how to optimally deploy the D-FACTS devices to maximize the MTD's effectiveness. Initial works on reactance perturbation-based MTD \cite{Morrow2012, Davis2012, RahmanMTD2014, LakshTPWRS2021} aim to use the pre-existing D-FACTS devices that are originally deployed to minimize transmission power loss and/or optimize system efficiency. These approaches do not consider relocating D-FACTS deployment or deploying new D-FACTS for the purpose of cybersecurity (i.e., implementing MTD). Subsequent works explored if MTD's effectiveness can be further enhanced by specifically designing D-FACTS deployment for this purpose \cite{ liu2020optimal, LiuBo2021}. 

MTD deployment strategies focus on minimizing the number of D-FACTS devices while maximizing effectiveness and reducing operational costs. \cite{liu2020optimal} uses the spanning tree method for D-FACTS deployment, measuring MTD effectiveness by the rank of the composite matrix. They select a minimum-weight spanning tree to minimize power losses while achieving optimal MTD effectiveness. \cite{LiuBo2021} proposed a strategy considering MTD's hiddenness, which often requires trading off some effectiveness. Their goal was to ensure hidden MTD solutions under any load conditions while maximizing effectiveness. This involved minimizing power flow differences and losses, and they used a depth-first-search algorithm to optimize D-FACTS deployment.

Beyond the spanning tree method, several works propose heuristic approaches for D-FACTS deployment strategies. \cite{ZhangMTD2020} uses the stealthy attack space as a metric, developing an iterative method that improves deployment until a preset goal, like minimizing the stealthy attack space, is achieved. Similarly, \cite{xu2021robust} evaluates MTD effectiveness based on the worst-case detection rate, iterating to reduce the undetectable attack space below a tolerance level. \cite{zhang2021strategic} formulates an optimization function to minimize vulnerable measurements, considering constraints like attack feasibility and D-FACTS limitations, and uses a heuristic method to explore near-optimal solutions. While these iterative methods are more computationally complex than the spanning tree method, they better accommodate practical constraints, such as limits on D-FACTS devices.

\subsubsection{Timing Aspects of MTD Rectance Perturbations}
Another key question is the frequency of MTD reactance perturbations. While attackers may not precisely detect MTD changes, they could infer them by monitoring sensor data and estimating the new topology. More frequent MTD operations enhance security but can disrupt system operations and increase costs. Two approaches are proposed for determining MTD frequency: (i) periodic and (ii) event-triggered perturbations. In the periodic approach, operators adjust reactances at regular intervals \cite{LakshTPWRS2021}, ideally before attackers gather enough information to launch undetectable attacks. Based on analyzing the attacker's learning process \cite{chin2017blind, LakshRMT2021}, reference \cite{LakshTPWRS2021} suggests that perturbation frequencies of a few hours are sufficient to invalidate attackers' knowledge. \cite{XuEventTriggered2023} and \cite{Higgins2021} propose an event-triggered MTD approach aimed at further reducing the frequency of MTD updates. We refer the reader to Section~\ref{MTD with machine learning} for more details.

\begin{table}[!t]
	\centering
    \resizebox{\columnwidth}{!}{
    	\begin{tabular}{|l|m{3.4cm}|m{3.4cm}|m{3.5cm}|}
		\hline
		Reference & Objective function & Specific constraints & Charateristics\\
        \hline
        \makecell[l]{\cite{Morrow2012,Davis2012} \\ \cite{RahmanMTD2014}} & - & - & Random perturbation \\
        \hline
        \cite{liu2020optimal} & Op. cost (losses) & Effectiveness (rank based) & \\
        \hline
        \cite{liu2020interior} & Op. cost (losses + OPF)  & Effectiveness (rank based) & Interior-point solver\\
        \hline
        \cite{LakshTPWRS2021} & Op. cost (OPF) & Effectiveness (SPA) & Cost-benefit tradeoff\\
        \hline
        \cite{TianHiddenMTD2019} & - & - & Hidden MTD\\
        \hline
        \cite{Zhang2020} & - & - & Hidden MTD + measurement protection\\
        \hline
        \cite{LiuBo2021} & Susceptance changes &  Hiddenness (Measurements) & Hidden MTD \\
        \hline
        \cite{Liu2022} & Effectiveness (BDD residual) & Hiddenness (BDD Residual) & Hidden MTD (AC)\\
        \hline
        \cite{CuiHidden2021} & Difference in measurements & Gradient of objective function & Hidden MTD (3-phase)\\
        \hline
        \cite{xu2021robust} & Effectiveness (SPA) &  Weakest points & MTD for noisy measurements\\
        \hline
	    \end{tabular}
     }
	\caption{The optimization formulations for existing MTD operation strategies.}
 \vspace{-5mm}
	\label{tbl:MTD operation}
\end{table}


\vspace{-3mm}
\subsection{Physics-based MTD Against CCPA} 
\vspace{-1 mm}
The majority of work in power grid MTD research literature is aimed at defending against FDI attacks. Alternatively, MTD can also be effective in defending against CCPAs \cite{LakshGT2021, LakshCCPA2019}. In contrast to FDI attacks that aim to only tamper with the measurement data, CCPAs consist of both a physical attack and tamper the measurements simultaneously (cyber-attack in this context). Physical attacks include causing line outages (e.g., by opening a circuit breaker) and coordinated cyber-attacks, such as an FDI against sensor measurements, mask the physical outages from being detected by the power grid's BDD. MTD to thwart CCPAs requires different criteria both in terms of D-FACTS placement and their operation.


Researchers in \cite{LakshCCPA2019, LakshGT2021} developed a framework for deploying and operating D-FACTS devices to thwart CCPAs. First, they introduced a D-FACTS deployment algorithm using the Maximum Weight Spanning Tree (MWST) approach. Then, they applied a game-theoretic method to determine which devices to perturb based on real-time conditions and assess likely attack targets. \cite{chen2022localization} extended this by localizing CCPAs through an MTD strategy and using ML-based detectors to identify the physical attack locations. \cite{zhang2022double} further refined this framework by addressing practical constraints like limited defense resources and focusing on protecting specific lines.

Follow up works have further aimed to improve MTD's performance against CCPA. For example, the researchers in \cite{yu2022moving} propose that, even after MTD activation, it may be still possible for an attacker to construct an undetectable CCPA if they can directly measure the phase angle difference between the two ends (buses) of the disconnected line. To address this vulnerability, they propose modifying the sensor measurement expression by adding a supplementary state factor, preventing potential attackers from obtaining the real values of system states, which are necessary for launching stealthy CCPAs. Similarly, researchers in \cite{hu2023controlled} employ controlled randomization on the set of measurements utilized in SE to mitigate CCPA, so that the attackers may not know the exact set of meters to target to launch undetectable CCPA. 


\vspace{-3.2 mm}
\subsection{MTD for Distribution Networks and Microgrids}
MTD strategies developed for transmission networks (detailed in Sections III-A and B) cannot be directly applied to distribution networks and microgrids. 
This is because distribution systems experience unbalanced loads, making the interactions between the phases more pronounced, and issues such as the impact on voltage stability need to be considered. Therefore, the MTD analysis must be based on detailed AC power flow and three-phase models. In micro grids, the X/R ratio for transmission lines is significantly smaller and reactance perturbation will result in a smaller overall impedance change, thereby reducing the effectiveness of D-FACTS operation in achieving MTD objectives \cite{LiuMCMTD2021}. 



In \cite{JhalaDistibution2021}, the authors developed MTD in distribution systems using voltage perturbation signals generated by inverter-based DERs. A detection mechanism is then implemented to check for the presence of the perturbation sequence in each sensor measurement, thereby identifying potential cyber attacks. The optimal set of DERs to generate the perturbation signal is determined using an optimization framework. \cite{Liu2018} and \cite{CuiHidden2021} propose reactance perturbation based hidden MTD strategy for balanced and unbalanced distribution networks. The MTD design is formulated as a nonlinear optimization problem using the full AC power flow model and solved using interior point and trust region based methods respectively.

The researchers in \cite{LiuMCMTD2021} and \cite{GiraldoMTDMicrogrid} design MTD strategy for microgrids. \cite{LiuMCMTD2021} proposed the use of a converter-based MTD, which shifts the primary control gains of the converter to strengthen the unknown input observer-based detector. They design the perturbations to improve the detection performance of the UIO-based detector while guaranteeing the voltage stability of the microgrid.  \cite{GiraldoMTDMicrogrid} studied a decentralized MTD scheme in microgrids. The main idea is to create random subsets of replicas of sensor/control data using IoT devices at a given time and combine them with a path selection algorithm that ensures that one of the selected replicas reaches the intended destination. This uncertainty created by these two actions is aimed at increasing the difficulty of learning the system details and corrupting the data shared between the sensor and controller.


\vspace{-3mm}
\section{{Network-based MTD}} \label{Network-based MTD}

\begin{figure}[!t]
\centering
\includegraphics[width=0.48\textwidth]{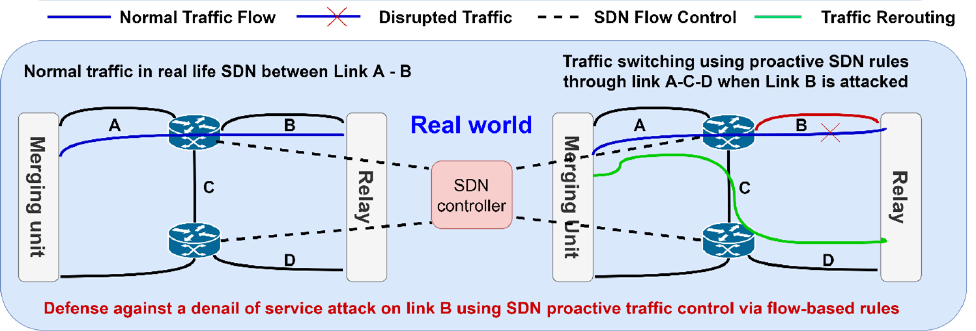}
\caption{Example of SDN-based cyber defense.}
\label{fig:SDN-based cyber defense}
\vspace{-6mm}
\end{figure}

MTD for strengthening network-based IDS has primarily focused on techniques such as IP-hopping or dynamically changing the communication path of the SCADA traffic in a software-defined SCADA network \cite{pappa2017moving, MTDSDN}. Specifically, the researchers in \cite{pappa2017moving} implement MTD using an IP-hopping strategy to dynamically and randomly mutate the IP address of the gateway router's external interface IPs, thereby preventing attackers from targeting the victim and vulnerability of the internal network. The researchers in \cite{MTDSDN} employed MTD strategies to safeguard power grid communication networks from Denial of Service (DoS) attacks. Their focus was specifically on enhancing the security of the SDN enabled Wide Area Network (WAN) in the context of power grid communications. The implementation of MTD involves manipulating the SDN control signals to periodically shift the communication channels. This proactive approach adds a layer of complexity for potential attackers, making it more challenging for them to identify the correct attack target. However, the literature only considers simple attacks such as single-point DoS attacks and lacks an in-depth analysis. 



MTD, which was originally developed for computer network security, has extended to protect cyber-physical systems \cite{chavez2019moving}. From a cyber perspective, one of the approaches is IP-Hopping in SCADA networks to dynamically change the IP addresses of different devices in the network \cite{pappa2017moving}. On the other hand, some efforts have focused on the development of theoretical foundations for different MTD strategies that particularly affect physical signals or physical connections to reveal stealthy attacks. Investigation of substation automation system with SDN is carried out in \cite{dong2015software}, which demonstrated that SDN performs better during fast failover, optimal bandwidth utilization, flow based match-action rules, and default security. SDN addresses the limitations of traditional OT networking for IEC 61850-based Substation Automation System (SAS) \cite{srivastava2024digital, yegorov2023analyzing, mustafa2023using}. Substation OT networks are generally ‘purpose engineered’, i.e., they serve a specific purpose on each segment of the network and do not change frequently compared to IT networks. This requirement allows for a more deterministic, predictable, and proactively engineered network design to support critical grid services, making SDN a suitable choice \cite{dehkordi2020retracted}. In \cite{MTDSDN}, SDN is employed as a MTD using a Mininet-based SDN simulation platform. In \cite{hussain2023teaching, mustafa2023cpgrid} authors used SDN in C37.118 based real-time testbed showing its advantages for synchrophasor networks. 

SDN employs network security by defining established flow-match traffic rules in contrast to traditional networking. To demonstrate, SDN-based defense, a cyber-physical system is modeled with software-defined networking (SDN) hardware, while the digital representation is modeled using NS3 \cite{hussain2023teaching, mustafa2023cpgrid}. The implementation of SDN has been achieved with a 2740 SDN hardware switch and a 5056 OpenFlow controller. The hardware component 2740S supports the IEC 61850 network requirements, while the hardware component 5056 allows the SDN switch management and configuration. Hardware 5056 also manages the network consisting of smart sensors, actuators, and real-time automation controllers.  On the other hand, NS3 is used for modelling nodes as 1-1 mapping with the substations. Cyber defence is enabled through MTD by network rerouting \cite{mustafa2023using}. Cyber defenses can be strengthened by SDN in various ways, such as faster reconfigurability and network re-convergence due to its capability of creating active and backup flows with priority.  Unlike traditional networking, SDN separates the control plane (that determines packet routing and direction) from the data plane (that carries the data), and a centralized controller is used to control the SDN switches using a protocol called OpenFlow. This separation allows the SDN switches to be only responsible for forwarding packets according to the rules made by the controller, which can be modified based on the type of attack/events.



Deception technologies, such as honeypot and decoy networks, can also contribute to the implementation of network-based MTD in a less intrusive way. Deception technologies in general offer dummy devices that behave like real devices (e.g., PLC and IEDs in smart grid systems). \cite{duan2018conceal} discusses further concepts of deception such as ``$k$-anonymity'', ``$l$-diversity'' and ``$m$-mutation''. $k$-anonymity offers a ``smokescreen'' by deploying $k-1$ identically-looking dummy devices for each real device to make it difficult for attackers to identify the real device as well as the topology. $l$-diversity further offers different configurations (e.g., different sets of services running) on the decoy devices to raise a bar for reconnaissance. $m$-mutation changes configurations, including network addresses, in a certain interval, realizing MTD. MTD using deception technologies can be implemented with no or minimal interference with real system infrastructure. For instance, as discussed in \cite{lin2020defrec}, messages sent by decoy devices can be ignored by real devices or SCADA HMI. Therefore, dynamic changes on these decoys do not affect the system operation, except for the increase of network traffic.  

While deception technologies for smart grids are not yet in a mature stage, we can find some efforts in the literature, such as \cite{lin2020defrec} and \cite{yang2020decied}. For instance, \cite{yang2020decied} aims at the implementation of scalable virtual IEDs to offer $k$-anonymity (i.e., an attacker would see $k$ identically looking devices) with different network addresses. Such internal state of such decoy devices is synchronized with the real device to imitate in a near real-time manner by means of multicast IEC 61850 GOOSE and SV messages sent by real devices and thus can behave in an indistinguishable way. One advantage of this approach is that no explicit or out-of-band communication among decoy devices is needed, and also deception does not require a back-end power system simulator to provide cyber-physical consistency, making the solution scalable. According to \cite{yang2020decied} a single industrial PC can run over 200 virtual IEDs, which can be further utilized for MTD, and such an industrial PC can be simply plugged into the station bus and process bus of the substation, alongside real IEDs. In addition to usage for network-based MTD, we also note that such decoy devices can emit fake, crafted power system measurements to implement other types of MTD strategies, such as physics-based ones, as discussed in \cite{yang2020decied}. This may help reduce the impact and required change on real cyber and physical devices. 

\vspace{-3mm}
\section{{MTD with Machine Learning}} \label{MTD with machine learning}
\vspace{-1mm}
Finally, we review the application of MTD with ML techniques, dividing them into two categories: (i) using ML to enhance MTD's defense capabilities, and (ii) applying MTD to make ML more resilient to adversarial attacks, as ML itself can be vulnerable.

In category (i), the researchers in \cite{XuEventTriggered2023} have introduced an event-triggered MTD approach, aiming to integrate ML-based IDS with physics-based MTD. The motivation stems from the fact that physics-based MTD incurs an operational cost, which can be excessive if implemented periodically. In the proposed scheme, the ML-based IDS serves as the primary detector, while the physics-based MTD serves as the secondary detector. In case the ML-based IDS raises an alarm, the data is further sent to an ML-based attack identification module that estimates the actual attack vector. Then, MTD is triggered to verify that the alarm indeed corresponds to an attack. The key idea is that following reactance perturbation if there is no attack, the subsequent measurements do not trigger the BDD, while in the presence of the attack, the BDD is triggered. 
The overall event-triggered approach is shown to significantly reduce the false alarm rate due to ML-based detectors and lower the operation frequency of physics-based MTD.


The researchers in \cite{chen2022localization} have integrated ML and physics-based MTD techniques to identify the location of physical attacks, specifically line outages, in CCPA. The motivation behind this integration lies in the limitation of physical IDS like BDD, which can only detect the presence of an attack. In contrast, ML-based IDS offers the potential to not only detect but also identify the location of the attack. However, the sophisticated CCPA can disrupt or circumvent the identification capabilities of ML-based IDS. To address this challenge, the researchers have implemented physics-based MTD to expose the stealthy CCPA, followed by the application of ML-based IDS to effectively localize the physical attack within CCPA. A model-agnostic meta-learning (MAML) is developed to enable the CNN to quickly adapt to the system reconfiguration \cite{chen2022meta}.



In category (ii), the MTD approach can be directly applied to ML models to enhance their robustness against adversarial attacks \cite{goodfellow2015explaining}. The application of MTD to strengthen ML-based detection is primarily proposed in the image processing domain  \cite{fMTD2019, sengupta2019mtdeep, Morphence2021}. MTD generates a diverse pool of ML models (e.g., neural networks used for attack detection), rather than a single ML model that is traditionally deployed in inference tasks. These diverse pools of models collaboratively make predictions and defend against attacks. The intuition is design the pool of ML models such that they present different defence landscapes toward adversarial attacks, while maintaining performance on essential prediction tasks. This is accomplished by reducing the transferability of adversarial attacks from between different models in the model pool. For example, the fMTD approach \cite{fMTD2019} generates diverse models by introducing random perturbations to a base model, finalizes predictions through majority voting, and periodically updates the model pool. MTDeep \cite{sengupta2019mtdeep} creates diverse models using various ML architectures (e.g., CNNs, MLPs, etc.). Morphence \cite{Morphence2021} generates diverse models using data transformation and adversarial training, with the final prediction determined by the most confident model. 


\vspace{-3mm}
\section{Conclusion and future work} \label{Conclusion and future work}
{MTD} is an emerging technique in power grids that aims to thwart sophisticated attacks against power grids.
In this paper, we have presented a comprehensive review of MTD in power grids. First, we present an overview of the different ways in which MTD is implemented in power grids. Then, we enlist the key performance metrics to measure the performance of MTD schemes and the trade-offs involved in the implementation of different MTD schemes. Then, we presented a detailed review of the MTD schemes of reactance-perturbation-based MTD to thwart FDI attacks, network-based MTD, physics-based MTD in microgrids and distribution networks, and  integration of ML with MTD schemes.

We conclude this paper by highlighting some open issues that remain in MTD design for power grids.
 (i) Unified MTD Design: Most of these works propose MTD design targeting a specific type of attack (e.g., FDI attack or CCPAs). However, the defender cannot know the type of attack that will be launched by the attacker apriori. Thus, moving forward, a unified MTD design that can potentially thwart multiple types of attacks will be important.
(ii) Combined cyber-physical MTD design: Currently, the research on MTD either targets \emph{moving} the components in the cyber layer (i.e., network-based MTD) or the components in the physical layer (e.g., reactance perturbation based MTD) etc. However, the research lacks joint consideration of cyber and physical aspects in the design that can further strengthen MTD.
(iii) MTD to strengthen ML models: Over the past decade, data-driven solutions based on ML algorithms have been increasingly adopted for detecting attacks against power grid \cite{sayghe2020survey, Musleh2020survey, Abdi2024survey, mohammadpourfard2021attack, lakshminarayana2022application, lakshminarayana2022data}. However, despite the effectiveness of such ML-based IDS, they have been shown to be vulnerable to \emph{adversarial attacks} \cite{goodfellow2015explaining, huang2022adversarial, AFDIA2021}.  
Thus, it is important to investigate methods for enhancing the resilience of ML-based systems against adversarial attacks. Current research literature lacks the design of MTD to defend against adversarial attacks targeting ML algorithms in the power grid, which must be addressed in the future. (iv) Effects of MTD on transient stability:  Existing literature primarily studies the effect of MTD perturbations on the power grid in terms of the steady-state metrics (e.g., increase in OPF cost or power losses). The effects of these perturbations on real systems in terms of voltage or frequency transients is still not researched adequately. (v) Real-world demonstrations: Despite the growing literature on MTD, most of these works analyze the design of MTD from a theoretical point of view (this is especially true for physics-based schemes such as reactance-perturbation-based MTD). Real-world demonstrations of the feasibility of introducing such perturbations in practical systems and the effects of device degradation are still lacking in the research community. Furthermore, a study on the perception of real-world grid operators on the feasibility of such perturbations also requires attention.


\vspace{-4mm}
\balance
\bibliographystyle{IEEEtran}
\bibliography{IEEEabrv,bibliography_reduced}

\end{document}